\documentstyle[12pt,epsf,epsfig,wrapfig]{article}  
\begin{document}  
\newtheorem{thm}{Theorem}
\newtheorem{cor}{Corollary}
\newtheorem{Def}{Definition}
\newtheorem{lem}{Lemma}
\begin{center}  
{\large \bf  
The spin-statistics theorem --- did Pauli get it right?
\ }  
\vspace{5mm}  

Paul O'Hara
\\  
\vspace{5mm}  
{\small\it  
Dept. of Mathematics, Northeastern Illinois University, 5500  
North St. Louis Avenue, Chicago, IL 60625-4699, USA. email:
pohara@neiu.edu \\}  
\end{center}  
\begin{abstract} In this article, we begin with a  review of
Pauli's version of the spin-statistics theorem and then show, by
re-defining the parameter associated with the Lie-Algebra structure of
angular momentum, that  another interpretation of the theorem may be given.
It will be found that the
vanishing commutator and anticommutator relationships can be  
associated with independent and dependent probability events respectively, 
and not spin value. Consequently, it gives a more intuitive understanding
of quantum field theory and it also suggests that the distinction between
timelike and spacelike events might be better described in terms of local and
non-local events.
Pacs: 3.65, 5.30, 3.70.+k    
\end{abstract}

\section {Introduction}

In Pauli's paper of 1940 \cite{pauli} the distinction between bose-einstein 
and fermi-dirac 
statistics is made according as to whether the commutator or anti-commutator 
relationships vanish in expression (20) of his paper. In modern terminology, we
would claim that bosons cannot be second quantized as fermions and vice-versa.
This can also be seen in modern formulations of the spin-statistics 
theorem where $[\psi_i(x), \psi_j(x^{\prime})]=0$ and $\{\psi_i(x), 
\psi_j(x^{\prime})\}
=0$ distinguish the two types of statistics, according as to whether they are
vector or spinor fields \cite{milonni}. 
Moreover, as direct calculation shows, 
once $\{\psi_i(x),\psi_j(x^{\prime})\}=0$ 
then $[\psi_i(x),\psi_j(x^{\prime})]\neq 0$ 
unless $\psi(x)_i\psi_j(x^{\prime})=0$. 

This theorem is then applied directly to particle systems by associating 
particles of integral spin with a vanishing commutator relationship and 
particles
of half-integral spin with the vanishing anticommutator relationship, from
which we conclude that all particles of integral spin are bosons and all
particles of half-integral spin are fermions. However on further investigation,
we realize that it is the identification of half-integer spin particles with 
the spinor representation of the rotation group that effectively forces
the distinction. In other words, if we were able to find a spinor represetation
for particles of integral spin then they too would have the characteristics 
of fermions, in accordance with the spin-statistics theorem.  

I would like to suggest that by re-defining the parameter of the Lie Algebra
associated with angular momentum, this in fact can be done. Moreover,
I would also like to suggest that this is justified for photons, since the 
expression ``spin angular momentum'' is a misnomer 
when applied to photons but serves rather to distinguish
two different polarized states.  Consequently, in this formalism, it is not 
spin-value that
determines whether the particles obey fermi-dirac or bose-einstein statistics
but rather the probability relationship between them. 

\section{Angular Momentum Theory}

In the usual development of angular momentum theory, we define
\begin{eqnarray} L^{\pm}=L_x\pm iL_y.
\end{eqnarray}
We then write
\begin{eqnarray}
L^2 &=& L^2_x + L^2_y + L^2_z\\
&=& L^{-}L^{+}+L^2_z+L_z,
\end{eqnarray}
from which it follows that
\begin{eqnarray} L^2\left|l\right>=l(l+1)\left|l\right>,
\end{eqnarray}
and that $\left|l\right>$ is an eigenvector of $L^2$.  
Similarly, the basis vectors
$\left|l\right>\dots \left|l-n\right>$ are eigenvectors of $L^2$ and $L_z$. 
Because of 
this, we denote the basis vectors of $L^2, L_z$ by $\left|l,m\right>$ with 
$-l\le m \le l$.
Now consider the operator  ${\bf L}=L_1+L_2$, where $L_1$
and $L_2$ are angular momentum operators as defined above. Denote the basis 
vectors of
${\bf L}^2, {\bf L}_z$ by $\left|l_1l_2LM\right>$,
where $|l_1-l_2|\le
L\le |l_1+l_2|$. In particular, when $l_1=l_2$
denote the joint state by $\left|llLM\right>$, then we can write\cite{brink} 
\begin{eqnarray} \left|llLM\right>=\frac12 \sum_{mm^{\prime}}
\bigl({\left|lm\right>_1
\left|lm^{\prime}\right>_2+(-1)^{L-2l}\left|lm^{\prime}\right>_1
\left|lm\right>_2}\bigr)\left<llmm^{\prime}|LM\right>.
\end{eqnarray}

In the case of electrons and deuterons, it is sufficient to assign respective
spin values of $1/2$ and $1$ in formula (5) above to obtain the correct 
probability distribution, associated with pairs of these particles. For example,
a calculation of the Clebsch-Gordan (C-G)coefficients for a pair of deuterons 
gives 
\begin{eqnarray}\left|2,0\right>=\sqrt{\frac{2}{3}}\left|0,0\right>+\frac{1}{\sqrt 6}
\left|1,-1\right>+\frac{1}{\sqrt 6}\left|-1,1\right>.\end{eqnarray}
On the other hand, if we assign a value of spin 1 to a photon, the above 
formula breaks down, since the probability of observing the  $\left|0,0\right>$
of two photons (given that $\left|\L=2, M=0\right>$ has occured) is 0 while
the C-G prediction would be $2/3$. Briefly put, the spin 0 state of a photon
is never observed.

How then can the above theory be extended to include spin 1 particles like 
the photon?  A simple remedy can be found provided we agree to re-define the 
angular momentum operator by $S_i=nL_i$ where $n$ is an integer, and define the 
subsequent Lie Algebra by
\begin{eqnarray} [S_i,S_j]=in\epsilon_{ijk}S_k.\end{eqnarray}
Note immediately that when $n=1$ we obtain the usual relationship for
spin $\frac 12$ particles, whereas for $n=2$ we obtain the usual 
properties for a spin 1 photon, and for $n=4$ we obtain the properties of the
spin 2 gravitons.  Using this modified algebraic structure, it is instructive 
to work out the theory of angular momentum in the usual way. We then quickly
find that
\begin{eqnarray}S^2\left|l\right>=n^2l(l+1)\left|l\right>=s(s+n)\left|l\right>.
\end{eqnarray}
Also $S^{\pm}S^z-S^zS^{\pm}=n^2(L^{\pm}L^z-L^zL^{\pm})=\mp n^2L^{\pm}=
\mp nS^{\pm}$ and it then follows that
\begin{eqnarray} S_zS^-\left|l\right>&=&n^2(L_-L_z\left|l\right>
-L_-\left|l\right>)\\
&=&n^2(l-1)L_-\left|l\right>\\
&=&(s-n)S_-\left|l\right>
\end{eqnarray}
Also by the usual arguments, there exists an r, such that 
$L^-\left|l-r\right>=0$ or equivalently $S^-\left|l-r\right>=0$. Hence
\begin{eqnarray}S^2\left|l-r\right>&=&(S^+S^-+S^2_z-nS_z)\left|l-r\right>\\
&=&(S^2_z-nS_z)\left|l-r\right>\\
&=&\{(s-nr)^2-n(s-nr)\}\left|l-r\right>
\end{eqnarray}
But $S^2\left|l-r\right>=s(s+n)\left|l-r\right>$ and therefore,
\begin{eqnarray}s(s+n)=(s-nr)^2-n(s-nr).\end{eqnarray}
Solving for $s$ gives $s=\frac{nr}{2}$. In particular when $n=2$, we obtain
the spin structure of a photon, with only two permissible states, which are
denoted by $(1,0)$ and $(0,1)$ respectively. Also, if we let $n=2$ in the 
commutator relations of equation (6), then the rotational properties of a 
polarized photon  can be modeled very effectively by the $SU(2)$ group with
parameter vector $2{\bf \theta}$, in contrast to similar properties of an 
electron 
(or neutrino) which are associated with the $SU(2)$ group with parameter 
${\bf \theta}$.

In other words, for photons and electrons the group representations are given by
$U_{2\theta}=e^{i{\vec \theta}.{\sigma}}$ and 
$U_{\theta}=e^{(i/2){\vec \theta}.{\sigma}}$ respectively, where ${\bf \sigma}$
represents the Pauli spin matrices. 
Moreover, an $SU(2)$ representation for the photon predicts a full 
angle formula, in contrast to the half-angle formula associated with the spin of
the electron (or neutrino). Furthermore, a C-G calculation based on the 
commutator relations (6) applied to photons (see next section),
naturally gives rise to the
triplet and singlet state decomposition associated with photons (without any 
need of a helicity argument), in contrast to the usual decomposition associated
with two spin 1 particles with an observable  spin-0 state. 

\section{Clebsch Gordan Coefficients for paired photons and deuterons}

Before reinterpreting the spin-statistics from the perspective of the 
generalized commutator relations of angular momentum, it is useful to
work out the respective joint states $\left|llSM\right>$ for pairs of photons 
and pairs of deuterons.

Consider two photons. Let $S=S_1+S_2$ represent their joint spins, and denote
their joint state by $\left|llSM\right>$. Then
three possible states emerge for $S=2$:
\begin{eqnarray}
\left|2,2\right>&=&\left|1,1\right>\\
\left|2,0\right>&=&\left<1,-1|2,0\right>\left|1,-1\right>+
\left<-1,1|2,0\right>\left|-1,1\right>\\
\left|2,-2\right>&=&\left|-1,-1\right>
\end{eqnarray}
which defines the triplet state. Also for $S=0$ we obtain
\begin{eqnarray}
\left|0,0\right>&=&\left<1,-1|0,0\right>\left|1,-1\right>+
\left<-1,1|0,0\right>\left|-1,1\right>
\end{eqnarray}
which defines the singlet state.
Note that the 
$$\left|2,0\right>=\left<1,-1|2,0\right>\left|1,-1\right>$$
state can be calculated directly by means of C-G coefficients, by observing that
$$\left<L,M_l|S^{\mp}S^{\pm}|L,M_L\right>=
4\left<L,M_l|L^{\mp}L^{\pm}|L,M_L\right>$$
from which it follows that
$$S^{\pm}\left|L,M_l\right>=2\hbar[(l\mp m)(l\pm m+1)]^{1/2}
\left|L,M_l\pm 2\right>.$$
In particular, if we set  $S^-=S^-_1+S^-_2$ then
$$2L^-\left|2,2\right>=2\hbar[(2+2)(2-2+1)]^{1/2}\left|2,0\right>$$
and 
$$\left|2,0\right>=\frac{1}{\sqrt2}\left|-1,1\right>+
\frac{1}{\sqrt2}\left|1,-1\right>.$$
Also if we assume orthogonality of the different states then 
$\left<2,0|0,0\right>=0$ implies 
\begin{eqnarray}
\left|0,0\right>&=&\frac{1}{\sqrt 2}\left|1,-1\right>-
\frac{1}{\sqrt 2}\left|-1,1\right>.
\end{eqnarray}

The deuteron is likewise a spin-1 particle. However in this case, the spin 0 
case can be observed. Moreover, a calculation of the C-G coefficients for a
pair of deuterons says a lot about the probability
weightings associated with the $\left|1\right>$, $\left|0\right>$,
$\left|-1\right>$ states of an individual deuteron. Direct calculation gives:
\begin{eqnarray}
\left|2,2\right>&=&\left|1,1\right>\\
\left|2,1\right>&=&\frac{1}{\sqrt 2}\left|1,0\right>+
\frac{1}{\sqrt 2}\left|0,1\right>\\
\left|2,0\right>&=&\sqrt{\frac{2}{3}}\left|0,0\right>+\frac{1}{\sqrt 6}
\left|1,-1\right>+\frac{1}{\sqrt 6}\left|-1,1\right>\\
\left|2,-1\right>&=&\frac{1}{\sqrt 2}\left|-1,0\right>+
\frac{1}{\sqrt 2}\left|0,-1\right>\\
\left|2,-2\right>&=&\left|-1,-1\right>.
\end{eqnarray}
It is worth asking, if the probabilities associated with the C-G coefficients
of the states can be calculated directly from conditional probability theory?
\footnote{Recall that for two events $A$ and $B$ defined on a finite sample 
space $S$, the conditional probability of $A$ given $B$ is denoted by $P(A|B)$
and $P(A|B)=P(A\cap B)/P(B)$ provided $P(B)\neq 0$.}
The answer is yes, provided the spectral distribution of an individual deuteron
has a probability distribution of 1/4, 1/2, 1/4 and not 1/3, 1/3, 1/3 which
is the current believe.
In particular, let $M_i$ where $i=1,2$ be a random variable associated with 
the spin of two independent deuterons such that
\begin{eqnarray}P(M_i=1)=P(M_i=0)=P(M_i=-1)=\frac{1}{3}\end{eqnarray}
Let $M=M_1+M_2$ be the sum of the spins. Note that $M$ too is a random variable
with values $2, 1, 0, -1, -2$. Then the conditional distribution
for the state $\left|2,0\right>$ associated with the two independent deuterons 
gives 
\begin{eqnarray*} P(M_1=0,M_2=0|M=0)&=&P(M_1=1,M_2=-1|M=0)\\
&=&P(M_1=-1,M_2=1|M=0)=\frac{1}{3}\end{eqnarray*}
However, this distribution is clearly different from the C-G calculations 
which gives 
$(\left<0,0|2,0\right>)^2=\frac{2}{3}$,
$(\left<1,-1|2,0\right>)^2=\frac{1}{6}$, 
$(\left<-1,1|2,0\right>)^2=\frac{1}{6}$. On the other hand, if  
\begin{eqnarray}P(M_i=1)=P(M_i=-1)=\frac{1}{4}\ \ \ 
P(M_i=0)=\frac{1}{2}\end{eqnarray}
then direct calculation shows that
\begin{eqnarray}&P&(M_1=0,M_2=0|M=0)=\frac{2}{3}\\
&P&(M_1=1,M_2=-1|M=0)=P(M_1=-1,M_2=1|M=0)=\frac{1}{6}\end{eqnarray}
which coincides with the C-G calculation.
Similarly, we find the remaining paired deuteron states are given by:
\begin{eqnarray}
\left|1,1\right>&=&\frac{1}{\sqrt 2}\left|1,0\right>-
\frac{1}{\sqrt 2}\left|0,1\right>\\
\left|1,0\right>&=&\frac{1}{\sqrt 2}\left|1,-1\right>-
\frac{1}{\sqrt 2}\left|-1,1\right>\\
\left|1,-1\right>&=&\frac{1}{\sqrt 2}\left|-1,0\right>-
\frac{1}{\sqrt 2}\left|0,-1\right>\\
\end{eqnarray}
and
\begin{eqnarray}
\left|0, 0\right>&=&\frac{1}{\sqrt 3}\left|1,-1\right>+
\frac{1}{\sqrt 3}\left|-1,1\right>-\frac{1}{\sqrt 3}\left|0,0\right>.
\end{eqnarray}
Note that $\left|0,0\right>$ case can be explained by assuming equation (26).

\section {A probability interpretation}

Thus far, we have given an algebraic interpretation of quantum angular 
momentum and also pointed out that by reparametrizing the $SU(2)$ group the 
spin value no longer serves to distinguish 
fermions from bosons.  However, a simple shift of emphasis from numerical value
to correlated and non-correlated states (particles), not only allow
fermions and bosons to be better understood in terms of non-local and local
events, or equivalently in terms of singlet and independent states but is also
very consistent with the usual quantum field theory formulation of the spin-
statistics theorem.

Indeed, we have already noted that the C-G calculations for two independent
deuterons, suggests that in general, the triplet state associated with
$S=n$ ($n=1$ for electrons and $n=2$ for photons)  
can be understood intuitively in terms
of two independent coin-flips. Moreover, it is completely symmetric under the 
exchange 
of spin values, with the states $\left|n,n\right>$, $\left|n,0\right>$ and
$\left|n,-n\right>$ having probability weightings 1/4, 1/2, 1/4 respectively.
In other words, not only do the states obey bose-einstein statistics, but
the above analysis suggests that such a statistics is a consequence of 
statistical independence. Moreover, since the
observation on one spin state has no influence on the observation of the
other state, by definition of independent probability, it further suggests
that any spin observable can always be
represented by $\sigma_z(x)=\sigma_z(x^{\prime})=\sigma_z$. It follows,
trivially, that $[\sigma_z(x),\sigma_z(x^{\prime})]=0$ and consequently,
statistical
independent states need to be quantized according to the commutator rules,
in keeping with a scalar quantum field.

On the other hand, in contrast to the triplet state, 
the singlet $\left|0,0\right>$ defines a rotationally invariant state and
obeys a fermi-dirac statistic. It is a pure state and cannot be 
decomposed, nor transformed into any other state without destroying the
rotational invariance.   
More specifically, consider two particles 
in the spin-singlet state:
$$\left|\psi\right>=
\left|+\right>\left|-\right>-\left|-\right>\left|+\right>.$$
and let $\vec \sigma(x)=(\sigma_1(x), \sigma_2(x), \sigma_3(x))$ and
$\vec \sigma(x^{\prime})=(\sigma_1(x^{\prime}), \sigma_2(x^{\prime}), 
\sigma_3(x^{\prime}))$ represent spin observables at $x$ and $x^{\prime}$
respectively.  
It then follows, because of the non-local interaction at $x$ and $x^{\prime}$, 
that the spin-singlet correlation permits the identification of the 
$\sigma_i(x)$ operator with the
$\sigma_i(x^{\prime})$ operator. Therefore,
$[\sigma_i(x), \sigma_j(x^{\prime})]=i\epsilon_{ijk}\sigma_k$ but 
$\{\sigma_i(x),\sigma_j(x^{\prime})\}=0$ and as a consequence,
non-local events in the form of spin-singlet states
need to be quantized according to the anticommutator rule. 
Moreover, $\sigma_i\sigma_j\neq 0$ and therefore events quantized according
to the anticommutator rule CANNOT be quantized with commutators (cf opening
paragraph of introduction), or equivalently, particles cannot be simultaneously
in non-local and local states; cannot be simultaneously subjected to the rules
of conditional and independent probability; cannot be simultaneously
rotationally invariant and non-invariant. However, once a measurement is
performed, the fermi-dirac state can be undone and changed into a bose-einstein
state; the singlet-state  can become two independent states.

\section {Conclusion}

The paper has tried to put together the various elements composing the 
spin-statistics theorem and has shown that the interpretations of Pauli's
original theorem can be generalized in a coherent way if the $SU(2)$ group is
parametrized in different ways, with Pauli's own theorem serving as
a special case.

In the process, we have also linked quantum statistics to classical
probability theory (cf equations (23), (28) and (29))
and made the prediction that a spectral decomposition of
a deuteron beam will have probability intensities 1/4, 1/2, 1/4 and not
1/3, 1/3, 1/3 which is the current quantum prediction. Hopefully, some
type of Stern-Gerlach experiment using deuteron atoms can be done to confirm
or negate this prediction. Moreover, it would seem that the prediction of 
1/4, 1/2, 1/4 is the
only one compatible with a C-G decomposition of angular momentum.

Finally, we note that the above analysis compliments the 
usual commutator and anticommutator relations of quantum field theory, 
associated with the scalar and Dirac fields respectively and gives a more
intuitive understanding  of these fields in terms of independent and dependent 
probability. It 
also further clarifies the analogy between the Pauli spin matrices and the
creation and annihilation operators of quantum field theory, 
referred to by Bjorken and Drell.\cite{bd}
Furthermore, this generalized spin-statistics theorem
gives a more natural and 
intuitive interpretation of microcausality. In the case of vanishing
commutators, microcausality implies that local events,
even at space-like differences, have simultaneous eigenstates and hence can
be in the same state,  in contrast to the microcausality associated with a
vanishing anticommutator which implies that two operators can never commute, 
can never have simultaneous eigenstates and hence can never be in the same 
state.
It then follows, from Pauli's (reparametrized) spin-statistics theorem that the 
spin-singlet state
obeys the fermi-dirac statistic and statistically independent spin states
obey the bose-einstein statistic. Moreover, this generalized result is in full
agreement with another formulation of the spin-statistics theorem to be found
in {\it Rotational invariance and the Pauli exclusion principle}.\cite{poh}

\end{document}